\documentclass[twocolumn,showpacs,amsmath,amssymb,superscriptaddress,aps,prc]{revtex4-1}

\usepackage[dvips]{graphicx}
\usepackage{epsfig}
\usepackage{amsmath}
\usepackage[dvips]{color}

\makeatletter

\newcommand{\Rmnum}[1]{\expandafter\@slowromancap\romannumeral #1@}
\makeatother

\begin{document}

\title{
Elliptic flow splitting between protons and antiprotons from hadronic potentials
}

\author{Pengcheng Li}
\affiliation{School of Nuclear Science and Technology, Lanzhou University, Lanzhou 730000, China}
\affiliation{School of Science, Huzhou University, Huzhou 313000, China}

\author{Yongjia Wang}
\affiliation{School of Science, Huzhou University, Huzhou 313000, China}

\author{Jan Steinheimer}
\affiliation{Frankfurt Institute for Advanced Studies, Ruth-Moufang-Str. 1, D-60438 Frankfurt am Main, Germany}

\author{Qingfeng Li}
\email{Corresponding author: liqf@zjhu.edu.cn}
\affiliation{School of Science, Huzhou University, Huzhou 313000, China}
\affiliation{Institute of Modern Physics, Chinese Academy of Sciences, Lanzhou 730000, China}

\author{Hongfei Zhang}
\affiliation{School of Nuclear Science and Technology, Lanzhou University, Lanzhou 730000, China}

\date{\today}

\begin{abstract}
The difference in elliptic flow $v_{2}$ between protons and antiprotons, produced in $^{197}\text{Au}+^{197}\text{Au}$ collisions at center-of-mass energies $\sqrt{s_{NN}}=5-12~\text{GeV}$, is studied within a modified version of the ultrarelativistic quantum molecular dynamics (UrQMD) model.
Two different model scenarios are compared: the cascade mode and the mean field mode which includes potential interactions for both formed and pre-formed hadrons.
The model results for the elliptic flow of protons and the relative $v_{2}$ difference between protons and antiprotons obtained from the mean field mode agree with the available experimental data, while the $v_{2}$ difference is near zero for the cascade mode.
Our results show that the elliptic flow splitting, observed for particles and antiparticles, can be explained by the inclusion of proper hadronic interactions.
In addition, the difference in $v_{2}$ between protons and antiprotons depends on the centrality and the rapidity window.
With smaller centrality and/or rapidity acceptance, the observed elliptic flow splitting is more sensitive to the beam energy, indicating a strong net baryon density dependence of the effect.
We propose to confirm this splitting at the upcoming experiments from Beam Energy Scan (BES) Phase-\Rmnum{2} at Relativistic Heavy Ion Collider (RHIC), the Compressed Baryonic Matter (CBM) at Facility for Antiproton and Ion Research (FAIR), High Intensity heavy ion Accelerator Facility (HIAF) and Nuclotron-based Ion Collider fAcility (NICA).
\end{abstract}
\maketitle
\section{Introduction}
\label{sec1}
The properties of Quantum Chromodynamics (QCD) and its phase transition from a quark gluon plasma (QGP) to hadronic matter is one of the main fields of research in heavy ion physics.
It is known from lattice QCD that for small baryon chemical potential ($\mu_{B}$), the thermodynamics properties of QCD becomes a smooth crossover \cite{PRD71034504,PRD85054503}.
As for QCD matter at large baryon chemical potentials ($\mu_{B}>400~\text{MeV}$), created in nuclear collisions at lower incident energies, various theoretical studies have suggested that the QCD phase transition may be of first order \cite{npa504668,plb442247,prc62054901,PRD78074507,ppnp72992013}.
Several experimental programs at the Brookhaven National Laboratory \cite{bes2}, European Organisation for Nuclear Research (CERN) and future Facility for Antiproton and Ion Research (FAIR) \cite{epja5360} and Nuclotron-based Ion Collider fAcility (NICA) \cite{NICA whitepaper} have been devoted to search for signals of the QCD critical point and phase transition and to study the properties of the QGP. One of the main goals of the Beam Energy Scan (BES) program at the Relativistic Heavy Ion Collider (RHIC) is to study the various aspects of the QCD phase diagram.
The BES Phase-\Rmnum{1} has been carried out at collision energies from $\sqrt{s_{NN}}= 200$ to 7.7 GeV, corresponding to baryon chemical potentials from 20 to 420 MeV.
BES-\Rmnum{1} has produced a large number of exciting results (see, e.g., \cite{PhysRevLett.120.062301,PhysRevC.93.014907,PhysRevLett.114.162301,STAR86054908,jpg38.124023,jpg38.124145}), and has restricted the region of interest to the collision energies below $\sqrt{s_{NN}}=20~\text{GeV}$.

A particularly interesting result of this experimental program was the high precision measurement of the elliptic flow of identified hadrons and their antiparticles, produced in the midrapidity region of the center-of-mass collision of the two heavy ions \cite{PhysRevC.88.014902}.
It was found that the elliptic flow of particles and their antiparticles shows a distinctive splitting which increases with decreasing beam energy or increasing net baryon density.

The anisotropic flow is an observable commonly used to study the properties of matter created in heavy ion collisions (HICs), sensitive to the equation of state (EoS) in the hot and dense early stage of the HICs. Various explanations have been proposed by several theoretical groups \cite{prc74064908,prc81044906,prc85041901,prc86044903,prc88064908,prl112012301,prd92114010,prc91024903x,prc91024914,prc91064901,EPJA52247} to account for the observed $v_{2}$ splitting of particles and corresponding antiparticles.
For instance, based on an extended multiphase transport model \cite{prc61067901,prc72064901} which includes mean field potentials in both the partonic and hadronic phase, the experimental data could be reproduced reasonably well at $\sqrt{s_{NN}}=7.7~\text{GeV}$ \cite{prc94054909}, if repulsive interactions for quarks were considered. In Ref.\cite{prc86044903}, within the framework of the ultrarelativistic quantum molecular dynamics (UrQMD) hybrid model, which combines the fluid dynamical evolution of the fireball with a transport treatment for the initial state and the final hadronic phase, the difference of elliptic flow is a result of the equilibrium hydrodynamical phase but is washed out by the subsequent transport phase through antiparticle annihilation. By tracing the number of initial quarks in the protons \cite{prc86044901}, it was proposed that the difference of $v_{2}$ between produced quarks and transported quarks may also contribute to the splitting \cite{CPC43054106}.

The purpose of this paper is to study the splitting of elliptic flow between protons and antiprotons at $\sqrt{s_{NN}}=5-12~\text{GeV}$ within the purely hadronic framework of the UrQMD model. Our goal is to show the effect of the well understood nuclear potentials of lower incident beam energies on the $v_{2}$ splitting at higher beam energies. Thus, providing an alternatively explanation which solely relies on lower energy (high density) physics, that only slowly subsides at higher beam energies. In the following we will show that also a purely hadronic description is able to reproduce the observed effect, if realistic hadronic interactions are taken into account. Predictions at even lower beam energies are made to check that indeed hadronic interactions are mainly responsible for the observed splitting.

\section{model}
\label{sec2}
In the following study the UrQMD model in its cascade (UrQMD/C, version 2.3) and a modified mean field mode (UrQMD/M) are employed. In order to accumulate a sufficient statistical accuracy, more than 10 million events are simulated for each energy and each mode. In the mean field mode of the UrQMD model \cite{ppnp41255,JPG251859}, hadrons are represented by Gaussian wave packets in phase space which read as:
\begin{equation}\label{Gau}
  \phi_{i}(\textbf{r},t)=\frac{1}{(2\pi L)^{3/4}} \text{exp}\left(-\frac{(\textbf{r}-\textbf{r}_{i})^{2}}{4L}\right)\text{exp}\left(\frac{i\textbf{p}_{i}\cdot\textbf{r} }{\hbar}\right),
\end{equation}
here $L=2~\text{fm}^{2}$ is the width parameter of the wave packet. The Wigner distribution function of the $i\text{th}$ hadron is represented as:
\begin{equation}
  f_{i}(\textbf{r},\textbf{p})=\frac{1}{(\pi\hbar)^{3}}\text{exp}\left(-\frac{(\textbf{r}-\textbf{r}_{i})^{2}}{2L}\right)\text{exp}\left(-\frac{(\textbf{p}-\textbf{p}_{i})^{2}\cdot2L}{\hbar^{2}}\right).
\end{equation}
The coordinate $\textbf{r}_i$ and momentum $\textbf{p}_i$ of hadron $i$ are propagated according to Hamilton's equation of motion:
\begin{eqnarray}
\dot{\textbf{r}}_{i}=\frac{\partial \langle H \rangle }{\partial\textbf{ p}_{i}},~~
\dot{\textbf{p}}_{i}=-\frac{\partial \langle H \rangle}{\partial \textbf{r}_{i}}.
\end{eqnarray}
The Hamiltonian $H$ consists of the kinetic energy $T$ and the effective interaction potential energy $U$, $H=T+U$.
The density distribution function of a single particle can be obtained from the Eq. (\ref{Gau}):
\begin{equation}
  \rho_{i}(\textbf{r},t)=\frac{1}{(4\pi L)^{3/2}}\text{exp}\left(-\frac{(\textbf{r}-\textbf{r}_{i})^2}{4L}\right).
\end{equation}

In order to study HICs at intermediate energies, a density- and momentum-dependent potential was implemented \cite{AICHELIN1991233,JPG32151}. In our study, both formed hadron and pre-formed hadron (string fragments that will be projected onto hadron states later) potentials are taken into account within the UrQMD model \cite{plb659525,plb623395,JPG36015111,mpla27,scpma59632001}.
The concept of pre-formed hadron can be found in the process of hadron production in deep inelastic scattering on nuclei \cite{NPA735277,NPA782224}.
In previous studies it was found that the description of observables such as nuclear stopping, elliptic flow and the Hanbury-Brown-Twiss interferometry (HBT) parameters of pions can be significantly improved by the inclusion of pre-formed hadron potential. Please see, e.g., \cite{plb659525,plb623395,JPG36015111,mpla27,scpma59632001} for more details.

For formed hadrons, density-, momentum-dependent and Coulomb potentials are included. The density-dependent potential can be written as:
\begin{equation}
  U=\alpha\left(\frac{\rho_{b}}{\rho_{0}}\right)+\beta\left(\frac{\rho_{b}}{\rho_{0}}\right)^{\gamma}.
\end{equation}
Where $\alpha=-268~\text{MeV}$, $\beta=345~\text{MeV}$, and $\gamma=1.167$ \cite{prc72064908}, corresponding to the nuclear incompressibility $K=314~\text{MeV}$.
The normal nuclear matter saturation density is given as $\rho_{0}=0.16$ fm$^{-3}$, and $\rho_{b}$ is the density of formed baryon and antibaryon.

The momentum-dependent term reads as:
\begin{equation}
  U_{md}=\sum_{k=1,2}\frac{t_{md}^{k}}{\rho_{0}}\int d\textbf{p}_{j}\frac{f(\textbf{r}_{i},\textbf{p}_{j})}{1+[(\textbf{p}_{i}-\textbf{p}_{j})/a_{md}^{k}]^{2}},
\end{equation}
where $t_{md}$ and $a_{md}$ are parameters, a detailed description
of the implementation can be found in \cite{prc72064908}.

For pre-formed hadrons, only a similar density dependent term as that for the formed baryons is used and the momentum-dependent term is neglected, which read as
\begin{equation}
  U=\alpha\left(\frac{\rho_{h}}{\rho_{0}}\right)+\beta\left(\frac{\rho_{h}}{\rho_{0}}\right)^{\gamma}.
\end{equation}
The parameters $\alpha$, $\beta$ and $\gamma$ are taken the same values as for formed baryons. $\rho_{h}=\sum_{i\neq j}c_{i}c_{j}\rho_{ij}$ is the hadronic density. For both formed and pre-formed baryons $c_{i,j}=1$, while $c_{i,j}=2/3$ for pre-formed mesons due to the difference of quark number, and 0 for formed mesons. Beside that, there are no interaction between pre-formed baryons and formed baryons. For formed mesons, no nuclear potential is considered. Since pre-formed hadrons are usually created in a dense environment, the interaction between these pre-formed hadrons will be mainly repulsive.

In the high energy region, the Lorentz-covariant treatment of the mean-field potentials is considered since the Lorentz contraction effect is significant. As done in \cite{PTP96263,prc72064908,prc97064913,EPJA5418}, the covariant prescription of the mean-field from the RQMD/S is implemented. RQMD/S uses much simpler and more practical time fixation constraints than the full RQMD \cite{RQMD192}, and can give almost the same results for the transverse flow as the original RQMD \cite{PTP96263}. The Hamiltonian, which reads \cite{prc72064908}
\begin{equation}
 H =\sum_{i=1}^{N}\sqrt{\mathbf{p}^{2}_{i} + m_i^2 + 2m_iV_i}.
\end{equation}
%
where $V_{i}$ is the effective potentials felt by the $i$th particle. And the equations of motion then become
\begin{eqnarray}
  \frac{d\mathbf{r}_{i}}{dt} &\approx& \frac{\partial H}{\partial \mathbf{p}_{i}}=\frac{\mathbf{p}_{i}}{p_i^0}+\sum_{j=1}^{N}\frac{m_{j}}{p_j^0}\frac{\partial V_{j}}{\partial\mathbf{p}_{i}}, \\
  \frac{d \mathbf{p}_{i}}{dt} &\approx& -\frac{\partial H}{\partial \mathbf{r}_{i}}=-\sum_{j=1}^{N}\frac{m_{j}}{p_j^0}\frac{\partial V_{j}}{\partial\mathbf{r}_{i}}.
\end{eqnarray}
The relative distance $\mathbf{r}_{ij}=\mathbf{r}_{i}-\mathbf{r}_{j}$ and $\mathbf{p}_{ij}=\mathbf{p}_{i}-\mathbf{p}_{j}$ in the two-body center-of-mass frame should be replaced by the squared four-vector distance with a Lorentz scalar,
\begin{eqnarray}
  \tilde{\mathbf{r}}_{ij}^{2} &=& \mathbf{r}_{ij}^{2}+\gamma_{ij}^{2}(\mathbf{r}_{ij}\cdot \mathbf{\beta}_{ij})^{2}, \\
  \tilde{\mathbf{p}}_{ij}^{2} &=& \mathbf{p}_{ij}^{2}-(p_i^0-p_j^0)^{2}+\gamma_{ij}^{2}\left(\frac{m_{i}^{2}-m_{j}^{2}}{p_i^0+p_j^0} \right)^{2},
\end{eqnarray}
where the velocity $\beta_{ij}$ and the $\gamma_{ij}$-factor between the $i$th and $j$th particles are defined as
\begin{equation}
  \beta_{ij}=\frac{\mathbf{p}_{i}+\mathbf{p}_{j}}{p_i^0+p_j^0},~~~~~\gamma_{ij}=\frac{1}{\sqrt{1-\mathbf{\beta}_{ij}^{2}}}.
\end{equation}

It is known that new hadronic degrees of freedom (hyperons, mesons, and quarks) are expected to appear in addition to nucleons in high densities, e.g., in HICs and in the neutron star interior. It has been argued that the different potentials for different particles can play an important role in the understanding of the EoS of neutron star matter \cite{14125722,APJ831,EPJA53121,PRSA474,epja5227}, as well as HICs matter. On another hand, for heavy-ion collisions in the energy range $\sqrt{s_{NN}}=5$ to 12 GeV studied in this work, the early dynamic processes is not so much determined by the Lambda-nucleon potential as nucleon-nucleon interaction, since the yield of Lambda is still small relative to that of the nucleon \cite{prl93022302,prc73044910,prc78034918,prc83014901,arxiv190800451}. Thus, it is expected that the inclusion of the Lambda potential would have a subleading contributions to the $v_2$ splitting between protons and anti-protons. Where the main contribution to the $\Lambda$ (and other hypeons) would also come from a reduced absorption, due to the lower density.

In the cascade mode of the model, no potential interactions are present and the hadrons interact only through binary scattering according to a geometrical interpretation of elastic and inelastic cross sections.

\section{results}
\label{sec3}
The anisotropic flow coefficients are defined by the Fourier decomposition of the azimuthal distribution of particles with respect to the reaction plane, which can be written as \cite{prc581671,PhysRevC.88.014902}:
\begin{equation}
  E\frac{d^{3}N}{d^{3}p}=\frac{1}{2\pi}\frac{d^{2}N}{p_{t}dp_{t}dy}\left[1+2\sum_{n=1}^{\infty}v_{n}\cos[n(\phi-\Psi_{RP})]\right],
\end{equation}
where $\phi$ is the azimuthal angle of the particles, $\Psi_{RP}$ is the azimuthal angle of the reaction plane, which is defined by the line joining the centers of colliding nuclei and the beam axis.
$\Psi_{RP}$ is fixed at zero in this work by definition. $v_{n}$ is the Fourier coefficient of harmonic $n$. The first harmonic coefficient of the Fourier expansion $v_1$ is called directed flow, the second coefficient $v_2$ is called elliptic flow and the third  $v_3$ is called triangular flow. In the following, we will discuss the elliptic flow ($v_2$), which defined as:
\begin{equation}
  v_{2}\equiv\langle \cos[2(\phi-\Psi_{RP})]\rangle=\left\langle\frac{p_{x}^{2}-p_{y}^{2}}{p_{x}^{2}+p_{y}^{2}}\right\rangle,
\end{equation}
here $p_{x}$ and $p_{y}$ are the two components of the transverse momentum $p_{t}=\sqrt{p_{x}^{2}+p_{y}^{2}}$. The angular brackets denote an average over all considered particles from all events.

For beam energies of ($1-11A$ GeV), the elliptic flow results from a competition between the early squeeze-out and the late-stage in-plane emission. The magnitude and the sign of the elliptic flow depend primarily on two factors \cite{E895831295,prl812438,science298}, the pressure of the compressed region and the passage time of the spectator matter.
At beam energies below $4A $ GeV, negative values for $v_{2}$ are observed \cite{E895831295,PLB612173,epja3031,NPA8761}, reflecting a preferential out-of-plane emission, as spectator matter is present and blocks the path of participant matter which try to escape from the fireball zone.
At higher energies the expansion occurs after the spectator matter has passed the compressed zone, and therefore the elliptic flow is mainly caused by the initial asymmetries in the geometry of the system produced in a non-central collision. An additional, very important, feature of increasing beam energy is the decreased amount of baryon stopping in the central rapidity region of the collision. At the highest beam energies available, at the Large Hadron Collider (LHC), one essentially observes a full particle antiparticle symmetry in all observables as no incoming baryons are stopped. Only as the beam energy decreases, more and more baryons are stopped in the experiments acceptance and therefore can have an effect on the difference between particles and antiparticles. Thus, it is natural to assume that any effect sensitive to the net baryon density will disappear slowly with increasing beam energy.

\begin{figure}[t]\centering
\includegraphics[width=0.5\textwidth]{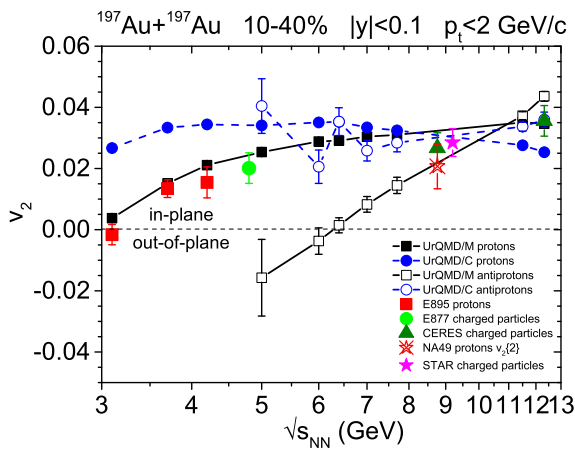}
\caption{(Color online) Collision energy dependence of the elliptic flow $v_{2}$ of protons and antiprotons in $^{197}{\rm Au}+^{197}{\rm Au}$ collisions in midcentral ($10-40\%$) collisions with $|{\rm y}|<0.1$ and $p_{t}<2~\text{GeV/c}$. The results are compared to data from different experiments for midcentral collisions. For E895 \cite{E895831295} and NA49 \cite{NA4968034903} there is the elliptic flow for protons. For E877 \cite{E877}, CERES \cite{CERES698253,0109017} and STAR \cite{STAR81024911} there is the $v_{2}$ for charged particles. Simulations with a pure cascade mode UrQMD/C (dashed lines with circles) are compared to results from UrQMD/M (solid lines with squares).}
\label{fig1}
\end{figure}

In our previous work, the elliptic flow of charged particles and protons for Pb+Pb collisions at Super Proton Synchrotron (SPS) energies was studied within the UrQMD model \cite{prc74064908}.
It was found that, the NA49 experimental data in the energy region below $E_{\text{beam}}=10A~\text{GeV}$ can be reasonably described by the UrQMD model with the inclusion of nuclear potentials. However, the model, based on the cascade mode, underpredicts the elliptic flow above $40A~\text{GeV}$. As an attempt to better describe the experimental data, the pre-formed particle potential was further considered \cite{plb659525,plb623395,JPG36015111,mpla27,scpma59632001}.
A strong repulsion is generated at an early stage with the inclusion of mean field potentials.
The repulsive interactions for the baryons make them expand faster, decreasing the local density and thus antiprotons have a smaller probability of being annihilated, which, furthermore enhanced the yield of antiprotons. Thus measurable yields of antiprotons can be quantitatively explained fairly well \cite{plb659525,mpla27}.
Also, the experimental data of protons and antiprotons elliptic flow in $^{197}{\rm Au}+^{197}{\rm Au}$ collisions at $\sqrt{s_{NN}}=7.7~\text{GeV}$ can be reasonably described \cite{scpma59632001}. In this work, we extend the model to describe the difference in $v_{2}$ between protons and antiprotons at $\sqrt{s_{NN}}=5-12~\text{GeV}$.

\begin{figure}[t]\centering
\includegraphics[width=0.5\textwidth]{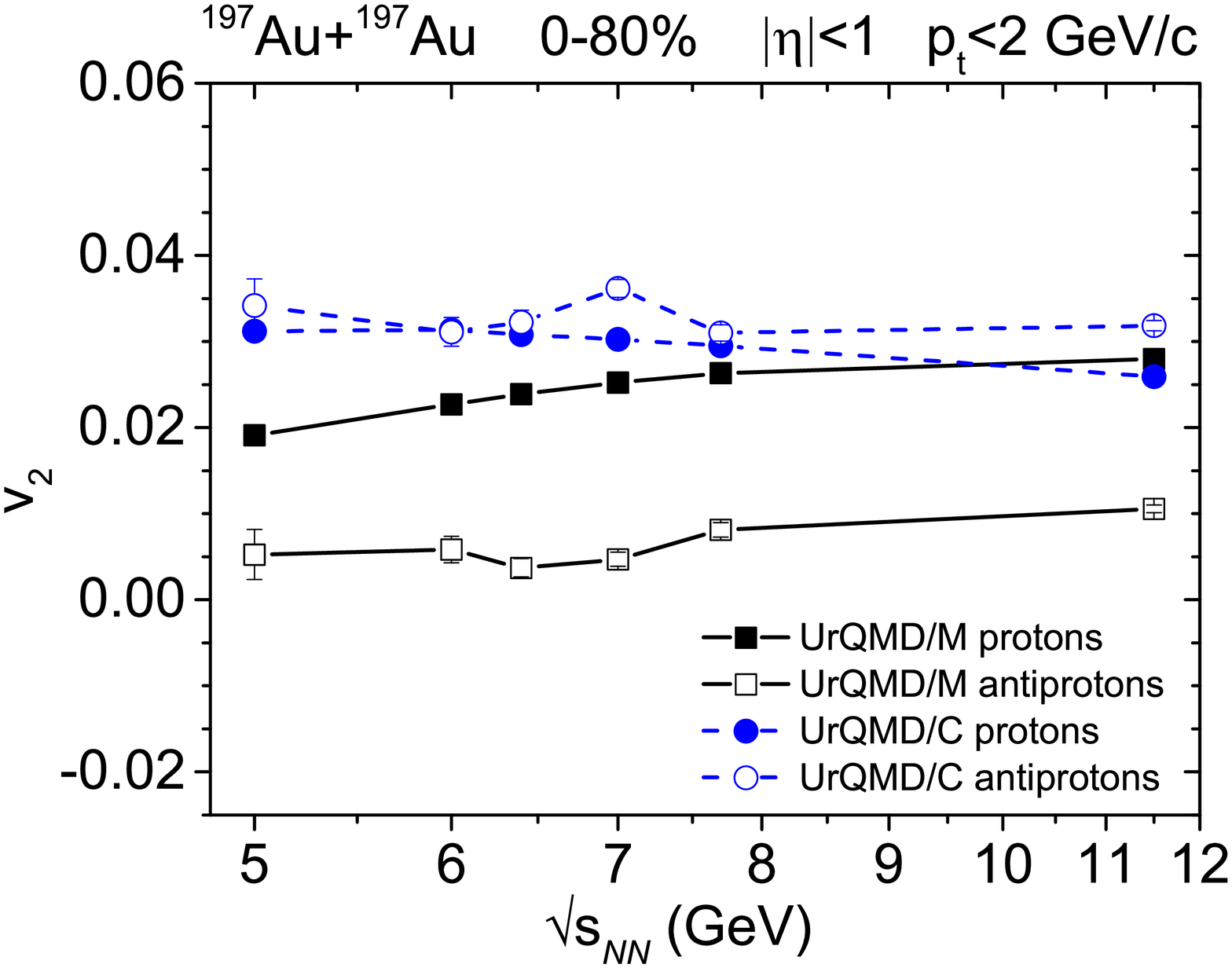}
\caption{(Color online) Elliptic flow of protons (line with solid symbols) and antiprotonss (line with open symbols) as a function of collision energy for $0-80\%$ central $^{197}{\rm Au}+^{197}{\rm Au}$ collisions with $p_{t}<2$ GeV/c and $|\eta|<1$.}
\label{fig2}
\end{figure}

Fig.\ref{fig1} depicts the UrQMD results for the energy dependence of the elliptic flow of protons and antiprotons for $10-40\%$ central $^{197}{\rm Au}+^{197}{\rm Au}$ collisions with $|{\rm y}|<0.1$ and $p_{t}<2~\text{GeV/c}$.
The calculated results from standard UrQMD cascade (UrQMD/C) mode and the UrQMD with mean field potential (UrQMD/M) mode are shown together with the experimental data from E895 \cite{E895831295}, NA49 \cite{NA4968034903}, E877 \cite{E877}, CERES \cite{CERES698253,0109017} and STAR Collaborations \cite{STAR81024911} for comparison.
As many experimental data as possible in this energy region are collected, although they are for various species of particles with different centrality and rapidity cuts.
The NA49 data \cite{NA4968034903} of protons from mid-central ($5.3-9.1~\text{fm}$) ${\rm Pb}+{\rm Pb}$ collisions is analysed by the cumulant method.
The STAR data \cite{STAR81024911} was taken with acceptance $p_{t}<2~\text{GeV/c}$, and averaged over pseudorapidity region $|\eta|<1$ from $0-60\%$ central Au+Au collisions. It is important to note that the STAR data are actually for charged particles and obtained using the event plane rather than the reaction plane which could lead to different results. Therefore this comparison should be taken with some caution.

For the elliptic flow of protons (line with solid symbols), the UrQMD/M mode which includes the pre-formed particle potential, is in line with experimental data, while the UrQMD/C mode overestimates the data at lower energies.
As for the $v_{2}$ of antiprotons (line with open symbols), the result from the UrQMD/M mode steadily increases as the energy increases.
Including nuclear potential, the difference in $v_{2}$ between protons and antiprotons, at midrapidity ($|{\rm y}|<0.1$) from $10-40\%$ central simulations, decreases with increasing beam energy, which is consistent with the trend of the SATR data \cite{PhysRevC.93.014907}. In the UrQMD/C pure cascade simulations, the $v_{2}$ values of antiprotons and that of protons are identical within errors.  As discussed earlier, in the UrQMD/M mode, both the $v_2$ of protons and anti-protons will be enhanced at the very early stage ($\sim$4 fm/$c$ for $\sqrt{s_{NN}}=7.7~\text{GeV}$, before most hadrons are formed), since a stronger early pressure is supplied by the potential of pre-formed particles.
With increasing time, the $v_{2}$ of protons is further increased by the repulsive potentials and a large number of two-body collisions.
Nevertheless, the $v_{2}$ of anti-protons keeps unchanged at the same period in time. Because most of anti-protons have been pushed out of the fireball and survived without a further annihilation process due to the decreased net-density, as well as potential modifications. And more detailed description of the early time dynamics, see our previous work \cite{scpma59632001}.
\begin{figure}[t]\centering
\includegraphics[width=0.5\textwidth]{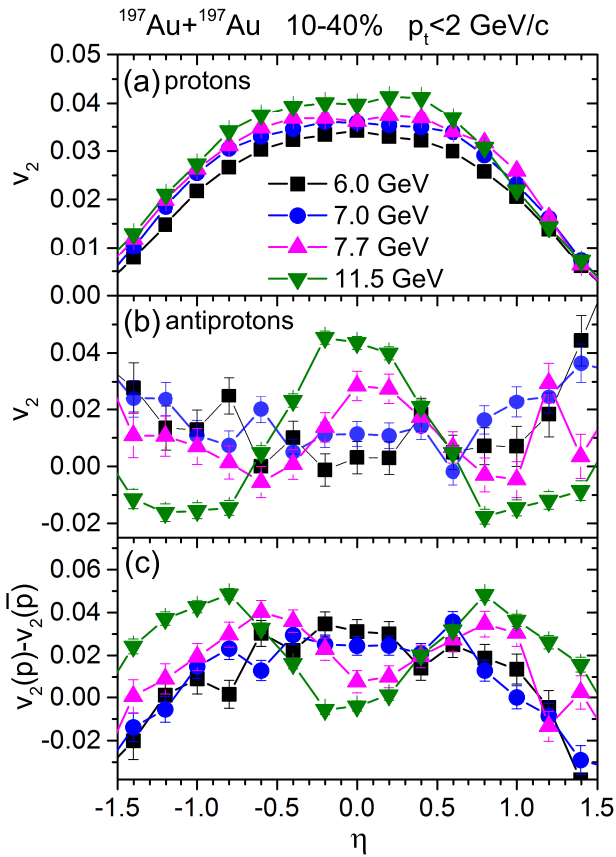}
\caption{(Color online) Pseudorapidity dependence of $v_{2}$ of protons and antiprotons as well as the absolute $v_{2}$ difference between protons and antiprotons for $\sqrt{s_{NN}}=6.0\sim11.5~\text{GeV}$ $10-40\%$ central Au+Au collisions. Only simulated with UrQMD/M model.}
\label{fig5}
\end{figure}

The influence of the acceptance windows on the energy dependence of the difference in $v_{2}$ between particles and corresponding antiparticles cannot be ignored.
In Ref. \cite{PhysRevC.93.014907}, the experimental data on the centrality dependence of the $v_{2}$ difference were presented.
The difference is larger for midcentral ($10-40\%$) collisions than for central ($0-10\%$) and peripheral ($40-80\%$) collisions.
In Fig.\ref{fig2} a larger centrality ($0-80\%$) and rapidity (here the so-called pseudorapidity $|\eta|<1$) bin is adopted. Again one can clearly observe that the elliptic flow of protons and antiprotons in the cascade simulation (UrQMD/C) are almost identical, which is similar to the result shown in Fig.\ref{fig1}.
However, the difference in $v_{2}$ between protons and antiprotons still exists in the simulations within the UrQMD/M mode, but the energy dependence is weaker, due to the larger centrality and rapidity windows. Taking a larger rapidity window decreases the effect for stopping, as even at higher beam energies regions with larger baryon density then fall into the acceptance. This finding supports the idea that the observed effect is mainly sensitive to the net baryon density of the system.

The influence of the rapidity cut on the difference in $v_{2}$ between protons and antiprotons is investigated further in Fig. \ref{fig5}.
The pseudorapidity ($\eta$) dependence of the elliptic flow $v_{2}$ of protons and antiprotons is respectively shown in panels (a) and (b) of Fig. \ref{fig5}.
The $v_{2}$ difference between protons and antiprotons is shown in the panel (c).
The elliptic flow of protons as a function of $\eta$ is qualitatively consistent with previous measurements \cite{NA4968034903,PRL94122303}.
At higher energy ($\sqrt{s_{NN}}>7.0~\text{GeV}$), the $v_{2}$ of antiprotons increases quicker than that of protons at midpseudorapidity.
And the difference in $v_{2}$ at midpseudorapidity decreases with increasing collision energy, mainly due to the $v_{2}$ of antiprotons at midpseudorapidity increases with increasing energy. However, it is opposite at the projectile and target pseudorapidity regions.
Thus in a wide pseudorapidity range $|\eta|<1$, the difference between $v_{2}$ of protons and antiprotons is only weakly dependent on the beam energy. As the net baryon density decreases in midrapidity the observed splitting effect slowly disappears.

\begin{figure}[t]\centering
\includegraphics[width=0.5\textwidth]{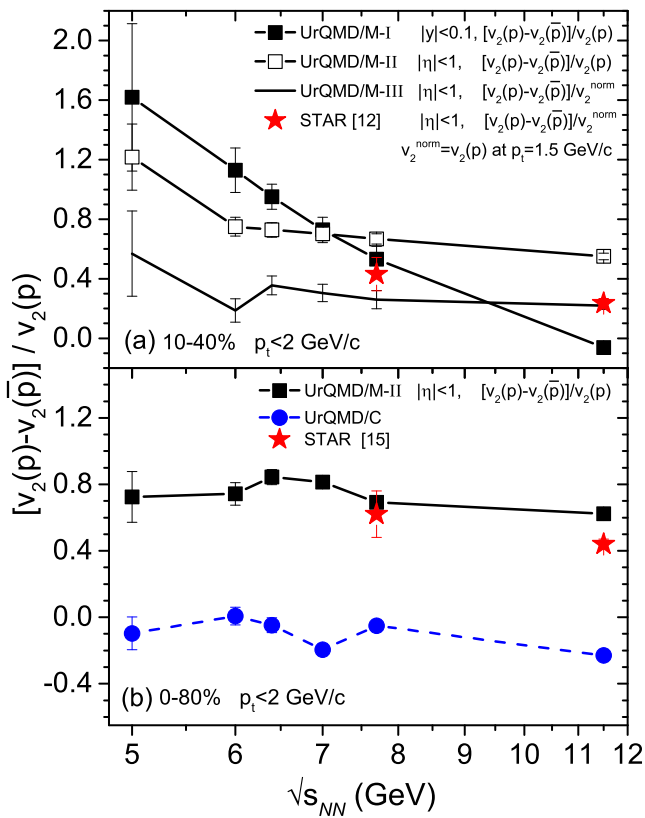}
\caption{(Color online) The relative $p_{t}$-integrated elliptic flow difference between particles and antiparticles versus the collision energy. The top panel is for $10-40\%$ central collisions which is simulated by UrQMD/M model. The bottom panel is for $0-80\%$ central collisions, calculated with UrQMD/C and UrQMD/M model. And the experimental data from STAR Collaboration is from Ref. \cite{PhysRevC.93.014907,jpg38.124023}.}
\label{fig3}
\end{figure}

The relative $p_{t}$-integrated elliptic flow difference between protons and antiprotons, defined by $[v_{2}(p)-v_{2}(\bar{p})]/v_{2}(p)$, is shown in Fig.\ref{fig3}, in which panel (a) is for $10-40\%$ central and panel (b) is for $0-80\%$ central Au+Au collisions, the experimental data taken from Ref. \cite{PhysRevC.93.014907} and Ref. \cite{jpg38.124023}, respectively.
In panel (a), the simulated results calculated from UrQMD/M mode are shown for three cases.
The UrQMD/M-\Rmnum{1} (solid line with solid squares) is the relative elliptic flow difference at midrapidity $|{\rm y}|<0.1$, while the UrQMD/M-\Rmnum{2} (solid line with open squares) is the relative $v_{2}$ difference at $|\eta|<1$.
The UrQMD/M-\Rmnum{3} (solid line) shows the relative $v_{2}$ difference at $|\eta|<1$ normalized by $v_{2}^{\text{norm}}$, the proton elliptic flow at $p_{t}=1.5~\text{GeV/c}$ as done in Ref. \cite{PhysRevC.93.014907}.
By comparing the results obtained with UrQMD/M-\Rmnum{1} and UrQMD/M-\Rmnum{2}, which differ only in the pseudorapidity window, we see that the relative $v_{2}$ difference decreases quickly with increasing energy at midrapidity. However, over a wider pseudorapidity range ($|\eta|<1$) this energy dependence becomes weaker, as discussed above.
When the same normalization as Ref. \cite{PhysRevC.93.014907} is employed in UrQMD/M-\Rmnum{3}, a weak energy dependence is observed, which is in qualitative agreement with the result of UrQMD/M-\Rmnum{2}.
In the simulations, the $v_{2}^{\text{norm}}$ is larger than the $p_{t}$-averaged $v_{2}$ of protons which used in UrQMD/M-\Rmnum{2}, thus the solid line is lower than the solid line with open squares.

In the panel (b) of Fig. \ref{fig3}, the relative $v_{2}$ difference at $|\eta|<1$ from $0-80\%$ central collisions within UrQMD/M and UrQMD/C model are shown.
In the UrQMD/C mode the relative $v_{2}$ difference is essentially zero.
The model containing potentials can quantitatively describe the STAR data \cite{jpg38.124023}.
At energies below $\sqrt{s_{NN}}= 7.7~\text{GeV}$ the difference seems to saturate and does not obviously depend on the beam energy.
By comparing the solid line with open squares in the panel (a) and the solid line with solid squares in the panel (b), both being the relative $v_{2}$ difference at $|\eta|<1$ but for different centralities, the energy dependence of the difference is gradually weakened with a larger range of the impact parameter.
We therefore suggest to measure the difference of $v_{2}$ between protons and antiprotons at various centralities and rapidity bins at lower beam energies as an indicator to explore the nuclear potential in this beam energy range. We note here that besides the $v_2$ splitting between protons and anti-protons, the $v_2$ splitting between other particles and anti-particles (e.g., $\Lambda$, $\Xi$, $\pi$, $K$) also have been measured by the STAR Collaboration \cite{STAR86054908,jpg38.124023}. The splitting for the $\Lambda$ seems similar to that of protons, indicating that the effect stems mainly from the bulk density and not so much from different potentials for different baryons.

\section{Summary and outlook}
\label{sec4}
To summarize, we have studied the elliptic flow of protons and antiprotons in heavy ion collisions at $\sqrt{s_{NN}}=5-12~\text{GeV}$, within the UrQMD model.
Two different modes of the model where employed: a pure cascade and the mean field mode.
The energy dependence of the elliptic flow of protons and the relative difference in elliptic flow $v_{2}$ between protons and antiprotons can be well reproduced within this purely hadronic description.
A stronger repulsion generated by the potential in the early stage leads to an earlier freeze out of the antiprotons.
The energy dependence of the $v_{2}$ difference between protons and antiprotons is gradually weakened with increasing the range of the impact parameter.
In addition, the difference in $v_{2}$ between protons and antiprotons in a narrow rapidity window is more sensitive to the beam energy variation than in a wide rapidity range. This indicates that the observed effect is strongly dependent on the net baryon number density.
Thus, the effect disappears and the nuclei become more transparent at higher beam energies. An interesting measurement would be the verification of this effect at large rapidities of higher beam energies.
In our simulations the elliptic flow splitting for $0-80\%$ central Au+Au collisions with $|\eta|<1$ still exists below the BES Phase-\Rmnum{1} energy region, and the splitting does not strongly depend on the collision energy. Meanwhile, we propose that the $v_2$ splitting of particles and anti-particles should be measured at existing and planned heavy ion experiments to highlight the importance of nuclear interactions even at the BES Phase-\Rmnum{2} ($\sqrt{s_{NN}}=7.7-19.6~\text{GeV}$) at RHIC, CBM ($\sqrt{s_{NN}}=2.7-4.9~\text{GeV}$) at FAIR and NICA ($\sqrt{s_{NN}}=4-11~\text{GeV}$).

Although our results explain the experimental data reasonably well, the difference in elliptic flow between particles and anti-particles might receive contributions from other effects, e.g., the different potentials for different particles \cite{prc85041901,prc94054909}, the chiral magnetic effect \cite{PRL107052303}, which are not included in the present work. However, these effects seem to be dominated by the nucleon potentials for all hadronic species. It is of particular interest to improve the present model by including these effects, to understand more deeply the energy-dependent difference in elliptic flow between particles and anti-particles for $\Lambda$, $\Xi$, $K$ and $\pi$. Only in such a study it can be truly understood whether the effect is due to the net-baryons density or also sensitive to different hadronic potentials.

Recently, we have noticed that in \cite{1907.03860v1}, J. Aichelin $et~al$ presents the novel microscopic n-body dynamical transport approach PHQMD (Parton-Hadron-Quantum-Molecular-Dynamics). They modified single-particle Wigner density $\tilde f$ of the the nucleon $i$ by accounts for the Lorentz contraction effect. It is more practical and time-saving to simulate the reaction in realistic heavy-ion calculation, and also provides reference and inspiration to our next investigations.

\section*{Acknowledgments}

This work has been supported by the National Natural Science Foundation of China under Grants No. 11875125, No. 11847315,  No. 11675066, the Zhejiang Provincial Natural Science Foundation of China under Grants No. LY19A050001, No. LY18A050002, and the ``Ten Thousand Talent Program" of Zhejiang province. The authors thank the computer facility of Huzhou University (C3S2). JS thanks the Samson AG and the Walter Greiner Gesellschaft zur F\"{o}rderung der physikalischen Grundlagenforschung e.V. for their support.

\end{document}